\documentclass[10pt]{article}

\usepackage{geometry}
\usepackage[parfill]{parskip}

\usepackage{graphicx}
\usepackage{epstopdf}

\usepackage{listings}
\usepackage{amssymb}
\usepackage{mathtools}

\usepackage{dsfont}

\usepackage{verbatim}

\usepackage{url}
\usepackage{hyperref}

\usepackage{color}

\usepackage{ifthen}

\usepackage[style=ieee,backend=bibtex,doi=false,isbn=false,date=year,dashed=false]{biblatex} 
\AtEveryBibitem{\clearfield{pagetotal}}
\usepackage{xpatch}
\xpatchbibdriver
{online}{\printtext[parens]{\usebibmacro{date}}}{\iffieldundef{year} {} {\printtext[parens]{\usebibmacro{date}}}}{}
{\typeout{There was an error patching biblatex-ieee (specifically, ieee.bbx's @online driver)}}

\newcommand{\todo}[1]{\ifthenelse{\equal{#1}{}}{{\color{blue} [TODO]}}{{\color{blue} [TODO #1]}}}

\title{SmartGridToolbox: A Library for Simulating Modern and Future Electricity Networks}
\author{Dan Gordon$^*$, Paul Scott$^*$, Sylvie Thiébaux$^*$ \\ $^*$The Australian National University}
\date{}

\begin{document}

\maketitle

\section*{Abstract}
We present SmartGridToolbox: a C++ library for simulating modern and future electricity networks. SmartGridToolbox is distinguished by the fact that it is a general purpose library (rather than an application), that emphasizes flexibility, extensibility, and ability to interface with a wide range of other tools, such as optimization technologies. It incorporates fully unbalanced network modelling, fast power flow and OPF solvers, a discrete-event simulation engine, and a component library that includes network components like lines, cables, transformers, ZIP loads and generators, renewable and storage components like PV generation and batteries, inverters, tap changers, PV, generic time dependent loads and more. We anticipate that SmartGridToolbox will be useful to researchers who require accurate simulations of electricity networks that go beyond simple applications of load flow - for example, by incorporating custom optimisation algorithms, controllers, devices, or network management strategies. Being a library, it is also perfect for developing a wide range of end use applications. We start with a comparison to existing open source software, and move on to present its main features and benchmark results. We conclude by discussing four applications, most notably, the use of SmartGridToolbox in the CONSORT Bruny Island Battery Trial, conducted between 2016 and 2019.

\section*{Keywords}
Open source software, electricity grids, modelling and simulation, power flow, load flow, OPF, DER, discrete event simulation.

\section{Introduction}
With the renewable energy revolution continuing to gather pace \cite{jones2017, weo2019, weo2020}, there is an urgent need for electricity networks to modernize their operational and regulatory frameworks, policy settings, physical and IT infrastructure and markets. A rapid increase in the uptake of EVs \cite{gev-outlook-2020} and household battery storage \cite{jones2017}
will make this even more pressing. There is therefore a clear case for governments, utilities, technology companies and researchers to have access to tools to evaluate these developments and to develop new ways and technologies for intelligently designing and operating electricity networks.

With this in mind, we have developed the C++ SmartGridToolbox library \cite{sgt} for modelling and simulating future and smart electricity grids. SmartGridToolbox incorporates fully unbalanced network modelling, power flow and OPF solvers, a discrete-event simulation engine, and a component library that includes network components like lines, cables, transformers, ZIP loads and generators, renewable and storage components like PV generation and batteries, inverters, tap changers, solar and weather modelling, generic time dependent loads and more.

SmartGridToolbox is distinguished by the fact that it is primarily a general purpose library (rather than an application) emphasizing flexibility, extensibility, and ability to interface with a wide range of other tools, such as optimization technologies, or existing code for devices. For example, in the CONSORT project \cite{franklin2016, scott2019}, we successfully interfaced SmartGridToolbox with distributed optimization software and commercial home energy system management software \cite{repositpower} running in the cloud. SmartGridToolbox was therefore able to act as a proxy for the real-life electricity grid during testing and what-if analysis of the NAC technology that was implemented on Bruny Island, Tasmania, Australia.
	
In this paper, we introduce and describe the SmartGridToolbox libraries. We describe two real-life field trials that have used SmartGridToolbox, and two other examples of SmartGridToolbox in action: the CONSORT project described above, the evolve project, a simulation of coordinated volt-VAR control using residential PV inverters, and a smart building simulation.

\section{Comparison to Other Software}
		
We concentrate here on open-source or free software, as it is of the most relevance to our interests in pure research and research and development. There are a number of open source and commercial packages available, that have some overlap with SmartGridToolbox. However, in our opinion, SmartGridToolbox occupies a unique niche, due to its flexibility and ability to interface with other software systems (being a C++ library), its sophisticated modelling of network components, its fast and efficient solvers and its inclusion of an event based simulation layer and component library.

GridLAB-D \cite{chassin2008, chassin2014, gld-website} is in many ways closest in concept and design to SmartGridToolbox. Like SmartGridToolbox, it includes a multi-agent, discrete event simulation engine coupled to detailed network modelling and various solvers. Compared to SmartGridToolbox, GridLAB-D presents a rich executable environment with many components and models, whereas SmartGridToolbox is a lighter-weight modern C++ library emphasizing optimization and permitting arbitrary customization and interfaceability with external algorithms and tools.

Like SmartGridToolbox, GridPACK \cite{gridpack-website} is a C++ library that could be adapted to a large number of end uses. Unlike SmartGridToolbox, it is aimed at a high-power-computing (HPC) context, with an emphasis on specialized parallel code, and less emphasis on network modelling and without the inclusion of an event driven simulation layer.

OpenDSS \cite{opendss-website} is a power systems simulation containing power flow solvers and a rich library of models of various traditional and smart grid components. It shares features with SmartGridToolbox and GridLAB-D, but it is written for Windows only in the Delphi language, and is thus somewhat limited in its application.

More geared toward traditional load flow studies, {\sc Matpower} \cite{zimmerman2020, zimmerman2011} is an open-source Matlab package that solves single-phase/balanced power flow and optimal power flow problems. PYPOWER \cite{pypower} ports {\sc Matpower} to python.

PowerModels \cite{coffrin2018} is written in the Julia language, using the JuMP optimization framework. It uses a similar data model to Matpower, but is more geared towards the solution of a variety of optimal power flow formulations using a several solvers.

Gravity \cite{hijazi2018} is a general-purpose C++ library for optimisation and machine learning that, like PowerModels, includes a variety of formulations for solving optimal power flow problems.

Pandapower \cite{thurner2018} is a more recent python library that extends PYPOWER. It is geared towards straight load flow studies, and doesn't include anything like the discrete event simulation found in SmartGridToolbox. Unbalanced modelling is limited to structurally balanced three phase networks with unbalanced loads, and the network modelling library is somewhat more generic. On the other hand, it includes features such as state estimation and short circuit analysis that are not currently part of SmartGridToolbox.

InterPSS \cite{zhou2017} is another open source package which applies distributed parallel computation to traditional load flow studies.

Modelica PowerSystems \cite{franke2014, modelica-ps-website} is based on the Modelica language \cite{modelica}, which brings a host of modelling tools and expressiveness, while at the same time limiting users to working within a specialized framework, and complicating and limiting the task of interfacing to external code.

Although not free or open-source, PowerWorld \cite{powerworld} is mentioned here, as it includes a freely-available educational version that is limited in the size of networks it can handle, but is suitable for educational purposes. It emphasizes diagrams and visualization. Similarly, PSS Sincal \cite{sincal} is a commercial package that is used in a range of contexts to model distribution networks.

\section{Architecture}
\begin{figure}[htb]
    \begin{center}
        \includegraphics[width=\columnwidth-4cm]{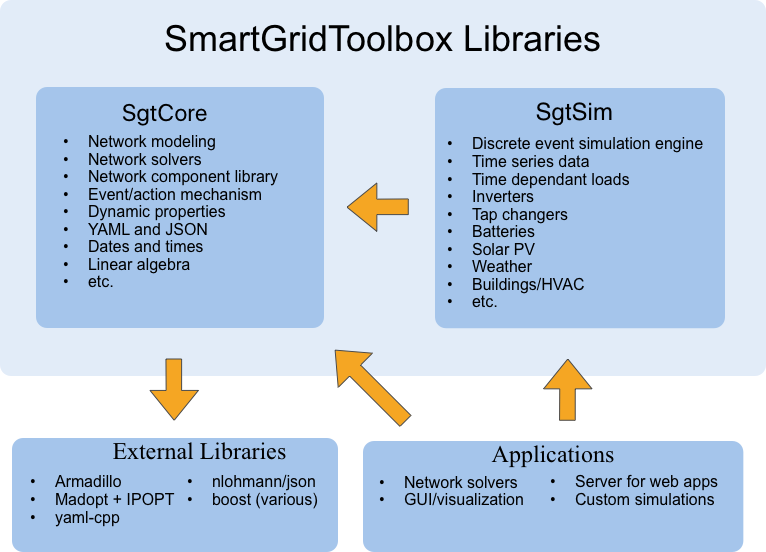}
    \end{center}
    \caption{
        SmartGridToolbox architecture. Arrows show code dependencies.
    }
    \label{FIG_ARCHITECTURE}
\end{figure}

SmartGridToolbox consists of two libraries, as shown in Fig. \ref{FIG_ARCHITECTURE}: SgtCore, which includes network modelling, power flow and OPF solvers, and various utility functions such as dates/times, linear algebra, parsers and so on, and SgtSim, which contains the classes and infrastructure of the discrete event simulation engine, as well as a library of simulation components. End uses that don't require discrete event simulation capabilities, or that may use a different simulation engine, are therefore able to use the core library without linking to SgtSim.

\section{The SgtCore Library}
The SgtCore library provides a framework for doing network modelling, power flow, and OPF. It can be used by itself, and also serves as a base for the SgtSim library. SgtCore also provides various utilities that are useful for both libraries, for example, an event mechanism, a mechanism for dynamic properties, the YAML parsing framework, and support for linear algebra, dates and times, time series, random number generation, and so on.

\subsection{Network modelling}
The SgtCore library provides a general framework for network modelling that is able to handle arbitrary phase configurations, including balanced/single phase networks and three phase networks, with and without a neutral wire.

\begin{figure}[htb]
    \begin{center}
        \includegraphics[width=\columnwidth-1cm]{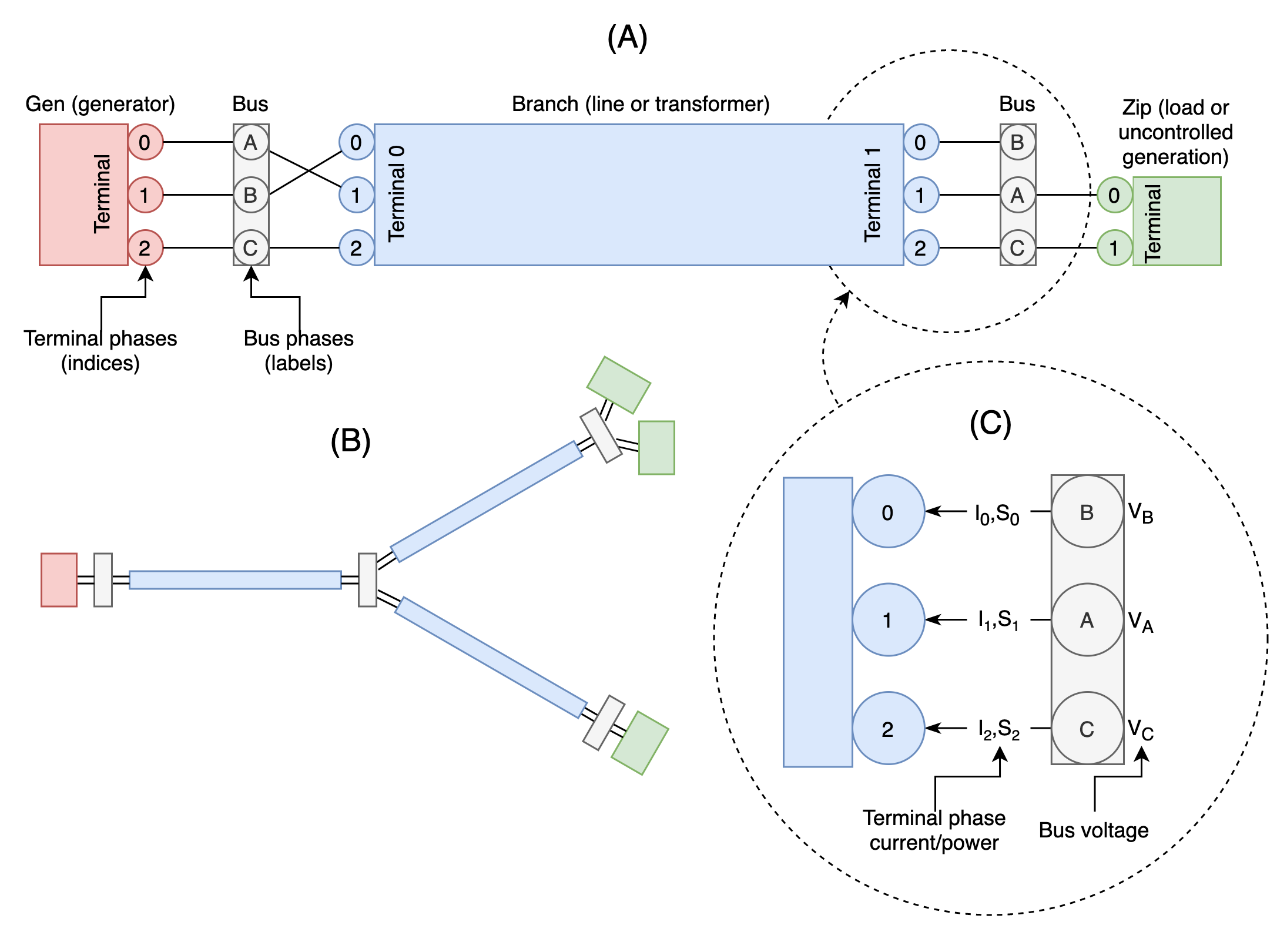}
    \end{center}
    \caption{
        The structure of a SmartGridToolbox network. (A) The manner in which branches, gens and zips are connected to buses, and the phasing of each connection. Each branch has two connection \emph{terminals}, and each gen or zip has a single terminal. (B) A whole network is built up from buses, branches, gens and zips connected in this way. (C) Buses have associated \emph{phase voltages}, and terminals have associated power and current flows, that run by convention from the bus into the terminal.
    }
    \label{FIG_SGT_NETWORK}
\end{figure}

Networks are modelled as a graph. The nodes represent \emph{buses}, comprising a set of co-located phases, each of which is characterized by a single voltage. The arcs represent \emph{branches}, which are typically either lines or transformers. Each branch has two terminals, labelled 0 and 1, each comprising a number of phases. Branch terminals connect to buses by mapping the terminal phases to a bus phases. Each bus can also have associated ZIP loads and/or generators, each with a single terminal that connects to a bus in the manner described above. The structure is shown in Fig. \ref{FIG_SGT_NETWORK}.

The network modelling framework contains the following classes and functions:
\begin{description}
    \item [Bus] A generic network bus, with arbitrary phasing.
    \item[Gen] A voltage-controlled generator, with information suitable for power flow or optimal power flow (OPF) solvers.
    \item[Zip] A load conforming to the ZIP model.
    \item[GenericBranch] A non-specific branch model, with given bus admittance matrix.
    \item[CommonBranch] A single phase Matpower style branch model: line plus transformer.
    \item[OverheadLine] An overhead transmission line with arbitrary phases.
    \item[UndergroundCable] An underground cable with arbitrary phases.
    \item[Transformer] A two winding transformer, with arbitrary phases and variable wiring. A wide range of single and three-phase transformers with various vector groups are possible. Taps and tap changers are modelled.
\end{description}

In addition to these classes, utility functions are available, for example, for generating approximate nodal admittance matrices using positive/negative and zero sequence components, applying the Kron reduction, and so on.

\subsection{Solvers}
SmartGridToolbox currently includes a Newton-Raphson AC-power flow solver and an extensible OPF solver that is based on the Madopt package \cite{madopt}.

\subsubsection{AC power flow equations}
Solvers are based on a generalized power flow model \cite{sgt-background} that allows ZIP loads between phases, for example, three phase delta-connected loads. Briefly, our power flow equations can be written in complex current injection form as:
\begin{align}
	\frac{S^*_{g,i}}{V^*_i} - \sum_{k \ne 0}{Y_{c,ik}V_{k}} - \sum_{k \ne i}\left[\frac{S^*_{c,ik}}{V^*_{ik}} + \frac{I_{c,ik}V_{ik}}{|V_{ik}|}\right] = 0
	\label{EQ_POWERFLOW_CPLX_I}
\end{align}
or in power injection form as
\begin{align}
	S_{g,i} - \sum_{k \ne 0}{Y^*_{c,ik}V_iV^*_{k}} - \sum_{k \ne i}\left[\frac{S_{c,ik}V_i}{V_{ik}} + \frac{I^*_{c,ik}V^*_{ik}V_i}{|V_{ik}|}\right] = 0
	\label{EQ_POWERFLOW_CPLX_S}
\end{align}

In these equations, the indices map to (bus, phase) pairs that physically represent a single piece of conductor with a single voltage. We will simply refer to them as buses below for simplicity. $V_i$ is the voltage at the $i$th bus,  and $V_{ik} := V_i - V_k$. $Y_{c,ik}$ is the nodal admittance matrix element $i, k$, $S_{g,i}$ is the complex power generation injection at bus $i$ and $S_{c,ik}$ $I_{c,ik}$ and $Y_{c,ik}$ are the constant power, constant current and constant impedance components of a ZIP load between buses $i$ and $k$.

In the AC power flow equations, the complex components of the $V_i$s are treated as variables, except possibly at special voltage controlled generator buses, where they may be treated as parameters, depending on the particular formulation. Similarly, the complex components of the generation $S_{g,i}$ may be treated as variables or parameters, depending on the type of voltage control. Other quantities are always treated as fixed parameters. Additional equations may be defined at generator buses to handle the voltage control.

\subsubsection{Newton-Raphson AC power flow solver} \label{SEC_NR}
SmartGridToolbox's Newton-Raphson solver is based mainly on the current injection method of Costa \emph{et al.} \cite{costa1999} and Garcia \emph{et al.} \cite{garcia2000}, but uses the generalized current-injection equations Eq. \ref{EQ_POWERFLOW_CPLX_I}, above. Due to the more complicated form of the generalized power flow equations, we find it convenient to work with the rectangular form of the Newton-Raphson equations. This means the voltage variables are the real and imaginary parts of the complex voltage, in contrast to polar equations that use the voltage magnitude and phase angle.

\subsubsection{AC OPF solver} \label{SEC_OPF_SOLVER}
SmartGridToolbox may optionally be built with an OPF solver, that makes use of the MADOPT \cite{madopt} library, which in turn relies on IPOPT \cite{wachter2006}. To formulate the OPF problem, we again start with the generalized power flow equations, this time in the power injection form, Eq. \ref{EQ_POWERFLOW_CPLX_S}. For OPF, we have found that using the polar form of the power flow equations leads to better convergence.

Equality constraints are defined with reference to the standard power flow equations, except that voltage constraints are relaxed at generator buses, so that the generation can vary within limits. Additional equality constraints are added at generators to ensure that the voltage magnitude is equal for all phases and that the voltage angle is maintained at the correct angle, e.g. a positive sequence set. Maximum power inequality constraints are added for branches. A cost of power is defined at generators, and this cost is minimized in the optimization.

The solver is written using template classes, so that additional user-defined variables and constraints may be added for buses, generators, branches and ZIP loads.

\subsubsection{Benchmarking} \label{SEC_BENCH}
\begin{figure}[!h]
    \begin{center}
        \includegraphics[width=\columnwidth]{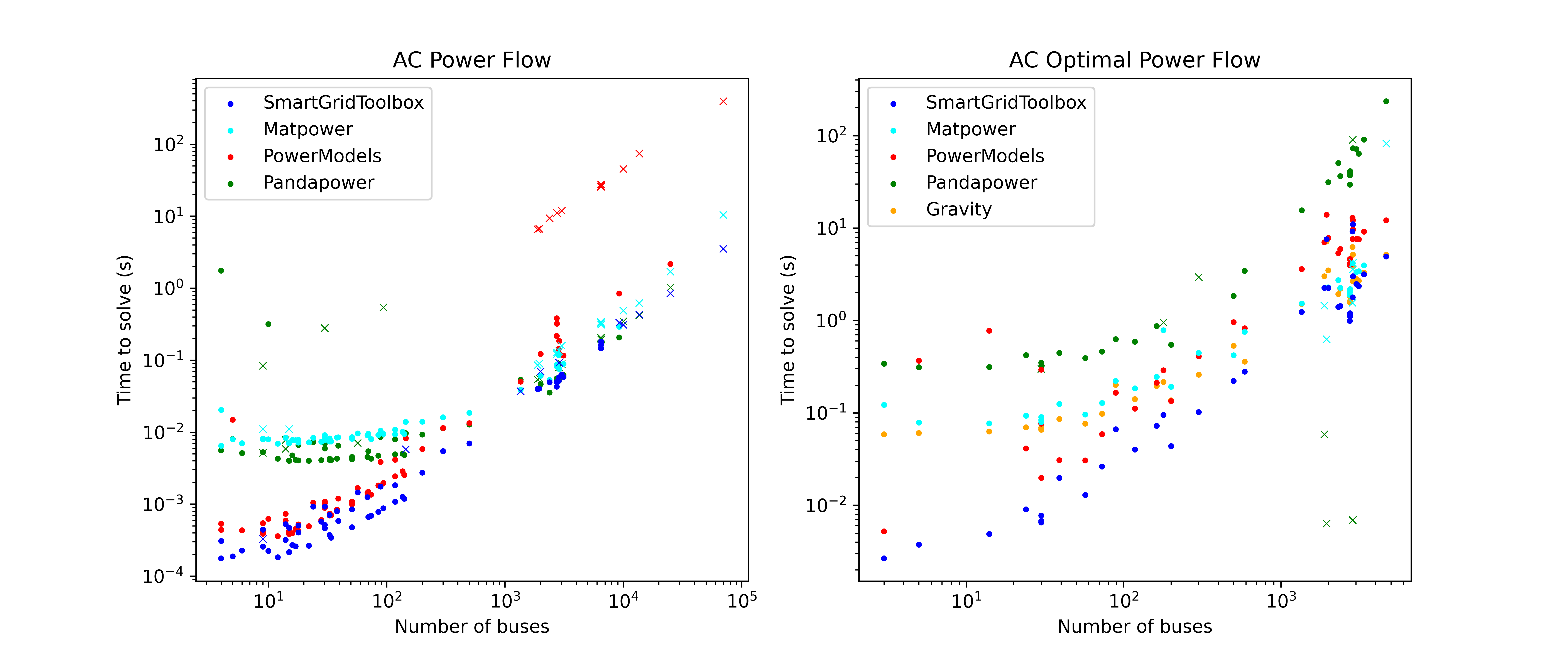}
    \end{center}
    \caption{Benchmarks for power flow (left) and OPF (right). Dots represent successfully solved cases, and `x's represent cases that failed to solve.}
    \label{FIG_BENCH}
\end{figure}

Fig. \ref{FIG_BENCH} shows the results of benchmarking SmartGridToolbox's power flow and OPF solvers against Matpower \cite{zimmerman2020, zimmerman2011}, Pandapower \cite{thurner2018}, PowerModels \cite{coffrin2018} and Gravity \cite{hijazi2018} (OPF only).

Power flow benchmarks use the Matpower 7.1 cases, while OPF benchmarks use the pglib-opf cases \cite{pglib}. The input files were modified to ensure that a flat-start was used in all cases. AC power flow and OPF were run without any special pre-solving\footnote{For example, Pandapower has the ability to use DC power flow to initialize variables for AC power flow, which significantly increases the range of convergence, particularly for larger problems.}.  SmartGridToolbox, PowerModels and Gravity were run using IPOPT with the HSL MA27 linear solver. PandaPower was run using Numba. Matpower was run using Octave. The benchmarks were run on an AMD Ryzen Threadripper 3990X Linux server with a clock speed of 2.9/4.3 GHz, 128 GB of RAM, and 64 cores.

I/O (input file loading) times were excluded from the results for solve time, but model-building times were included\footnote{We included model-building times for two reasons: first, model-building is an unavoidable overhead, and second, we found it hard or impossible to separate model building time from model solving time without modifying code}.

The final solved objective function values of the SmartGridToolbox OPF results were checked against Matpower results and were found to agree in all cases.

For AC power flow, SmartGridToolbox is significantly faster than the other packages for fewer than 2000 buses, after which the advantage becomes less marked. Failed runs are shown using `x's; in these cases, the runtime simply reflects how hard the solver tries before giving up, and is therefore not particularly relevant. We speculate that for more than 2000 buses, the backend linear solver, as opposed to other data processing and language constraints, becomes the main bottleneck, and that SmartGridToolbox has less of an advantage here.

For AC OPF, the results follow a similar pattern to the AC power flow benchmarks, with SmartGridToolbox being generally faster below about 2000 buses, and thereafter showing similar performance to several of the other packages.

It is also apparent that AC power flow is more than an order of magnitude faster than OPF. This underscores the advantage of using a good custom Newton-Raphson solver where applicable.

\subsection{YAML configuration files and parsing}
SmartGridToolbox provides a framework for parsing YAML configuration files. A \verb|Parser<T>| template is provided, whose function is to modify an object of type \verb|T| based on the contents of a YAML document. For example, a \verb|Parser<Simulation>| object is able to parse a YAML configuration document, and make resulting changes to a \verb|Simulation| object, e.g. by populating the simulation with parsed-in \verb|SimComponent|s, setting start and end times, and so on.

Parsers may register one or more \verb|ParserPlugin|s. A SmartGridToolbox YAML configuration file consists of a list of key-value pairs, with most of the keys corresponding to a specific subclass of \verb|ParserPlugin|. For example, take the following YAML fragment:
\begin{verbatim}
            - solar_pv: {id: solar_pv, efficiency: 0.25, area_m2: 120 ...}
\end{verbatim}
When the ``solar\_pv'' keyword is encountered, the YAML sub-object \texttt{\{id: solar\_pv, ...\}} is passed to a \verb|SolarPvParserPlugin| object for parsing. By writing new ParserPlugins, the user can enable their own custom objects to be parsed in as part of the YAML configuration file.

\subsection{Properties}
While C++ has many advantages such as speed and universality, it is based on a static type system that may be limiting or inconvenient for extensible component-based applications like SmartGridToolbox. For example, in a GUI (see Section \ref{SEC_VIEWER}), we would like to be able to discover, get, and set the properties of an object during runtime, so that we don't need to write code to inform the GUI about the properties each new custom class that we write. To meet these needs, SmartGridToolbox provides a special mechanism for defining runtime properties of arbitrary type, that are inherited just like class members. Properties may be accessed either as JSON values (which is convenient for interfacing with many applications), or as their particular C++ type. 

\subsection{Component collections}
SmartGridToolbox makes extensive use of a \verb|Component| base class, for both network modelling (where \verb|Component| is a base class for network components such as buses, lines and loads), and in discrete event simulations, where it is a base class for simulation objects such as time varying loads, solar PV, batteries, and so on. \verb|Component|s are essentially just objects with a string ID and type.

To store and reference collections of arbitrary objects (normally, but not necessarily, \verb|Component|s), a \verb|ComponentCollection<T>| class template is provided. \verb|ComponentCollection|s combine features of \verb|std::map| and \verb|std::vector|. They preserve order of insertion (which is, for example, important for preserving the order of updates in discrete event simulations), and components may be retrieved either by index or by key. References to members of a \verb|ComponentCollection| will remain valid on replacement of the referenced element, with the reference pointing to the newly replaced element. 

\subsection{Other utility classes}
Complex numbers are \verb|std::complex<double>|. Time is represented internally by the various classes from the \verb|boost| time libraries, with various utility functions for outputting, parsing, converting to posix timestamps, and so on. Vectors, matrices, and linear algebra are handled by the Armadillo library \cite{sanderson2016, sanderson2018}. Various logging and error reporting functions are provided. JSON support is handled by the \verb|nlohmann::json| library \cite{Lohmann_JSON_for_Modern}. SmartGridToolbox makes extensive use of the YAML data format \cite{yaml}, for which purpose the yaml-cpp library is used \cite{yaml-cpp}. Random number generators from the boost random library \cite{boost-random} are also provided.

\section{The SgtSim Library}
SmartGridToolbox simulations employ agent-based and discrete event simulation concepts. Simulation objects are modular entities, largely responsible for their own evolution in time, and interacting with other objects in a structured manner. To handle the range of timescales that may exist within a single simulation, objects are responsible for signalling their own discrete updates, at whatever timescale is appropriate to their evolution. These updates may entail further contingent updates and structured interactions with other simulation objects.

SmartGridToolbox simulations proceed according to the following logic:
\begin{enumerate}
    \item \label{IT_RANK}
        An evaluation order ranking for simultaneous component updates is  calculated, based on specified dependencies.
    \item
        Components are initialized to their starting state.
    \item
        In a second initialization pass, components are given the opportunity to store references to other components and to set up dependencies.
    \item \label{SIM_STEP_TIMESTEP}
        Each component provides to the \verb|Simulation| a time at which it is next scheduled to update its state. Such updates are termed \emph{scheduled updates}. The current timestep is set to the earliest update time of all components.
    \item
        Components that have a scheduled update at the current timestep are processed, in order of their rank, see item \ref{IT_RANK}. As a consequence of an update, other components may be added to a list of \emph{contingent updates} at the given timestep. This means that they will be asked to update their state during the timestep, regardless of whether they have a scheduled update. When a component undergoes a scheduled update, it is removed from the list of contingent updates. The logic is that it must have been placed on the list by a component that updated prior to it, so causality will be preserved by the scheduled update.
    \item
        When all scheduled updates in a timestep have occurred, the contingent updates are processed. This \emph{two phase} process \cite{pidd1998} minimizes unnecessary updates by ensuring that, if the component is asked to undergo a contingent update by more than one other scheduled component, it will only need to update once after the updates of all these components.
    \item
        When there are no more contingent updates, the timestep is deemed to be complete, and the next timestep is started at step \ref{SIM_STEP_TIMESTEP}.
\end{enumerate}

\subsection{Event-action mechanism}
For discrete event simulations, components should be able to communicate in a sophisticated manner both actively, by pushing updates on their known dependents, as well as passively, by observing the behaviour of other components and then triggering actions that occur as a result of certain conditions. To enable this kind of interaction, we have implemented a variant of the observer pattern, that we term an \emph{event-action} architecture\footnote{In fact, the event-action code is part of the SgtCore library rather than SgtSim, as it has wider applicability beyond discrete event simulations. However, we discuss it in the section on Simulation, as it is most relevant to this material.}. Each component may have a number of \emph{events} associated with it, that are triggered whenever certain conditions are met. For example, a battery may trigger an event when its charging or discharging power changes, for example when it runs out of charge. Event observers (e.g. an inverter to which the battery is connected) may store an arbitrary set of actions that they wish to be performed when the event is triggered.

\subsection{Network simulation}
The time dependent behaviour of networks is simulated in the quasi-steady-state approximation, i.e. by solving the power flow equations at each timestep. This ignores the sub-second and microsecond transient dynamics that may occur due to to rotor dynamics in generators, the waveguide-like properties of lines, and frequency dependence in other components. In principle, it is a straightforward matter to embed fast timescale solvers operating in a standard timestepped manner within the umbrella of the discrete event simulation architecture. Although we have ambitions for transient solver featuring rotor dynamics (the swing equations), this is not currently part of the current SmartGridToolbox distribution.		

\subsection{Simulation component library}
The simulation library also contains a library of \verb|SimComponents|.

Network simulation components (buses, branches, generators and loads) can be handled by lightweight classes (\verb|SimBus|, \verb|SimBranch| etc.), that wrap the core network component wrapper, taking care of dependencies and updates. A \verb|SimNetwork| wraps the core network, making sure that the network is appropriately updated when any of its constituents register a change of state.

Transformer tap changers are included in the form of \verb|AutoTapChanger| and \verb|TimeSeriesTapChanger| classes. The former extends the core tap changer functionality by including tap changer delays and handling the dependencies that exist between tap changers and the network. The latter allows historical tap changing data to be replayed during a simulation.

A \verb|TimeSeriesZip| load class is included to handle all kinds ZIP loads and non-dispatchable generators whose time dependence can be modelled using time series data.

\verb|Battery|s and \verb|SolarPv| classes are included, and an \verb|Inverter| class is provided to handle these DC power sources/sinks.

The output of a \verb|SolarPv| object depends on a \verb|Weather| object. The \verb|Weather| class handles aspects of the weather such as temperature, cloud cover and solar insolation.

A simple thermal model of a building is also included; it too depends on a \verb|Weather| object.

The \verb|Heartbeat| class can be used to force regular updates.

The \verb|RealTimeClock| class can induce updates in real time; the latter was used in the Bruny Island Battery Trial (see section \ref{SEC_CONSORT}) to test the software components that were eventually deployed in a live field trial.

It is part of the SmartGridToolbox philosophy that users should be open to writing their own custom \verb|SimComponent|s, particularly for controller components. In the near to medium future, we hope to continue adding various other components, such as different sources of generation etc. to the SgtSim library.

\section{Other Extensions}
\subsection{Python bindings}
Python bindings have been written for the SgtCore library (with plans for some of the SgtSim functionality to be added). These cover much of the functionality of the C++-library, and have been used by the evolve DER Project \cite{evolve-arena}, funded by the Australian Renewable Energy National Authority (ARENA).

\subsection{Visualization} \label{SEC_VIEWER}
\begin{figure}[htb]
    \begin{center}
        \includegraphics[width=\columnwidth-6cm]{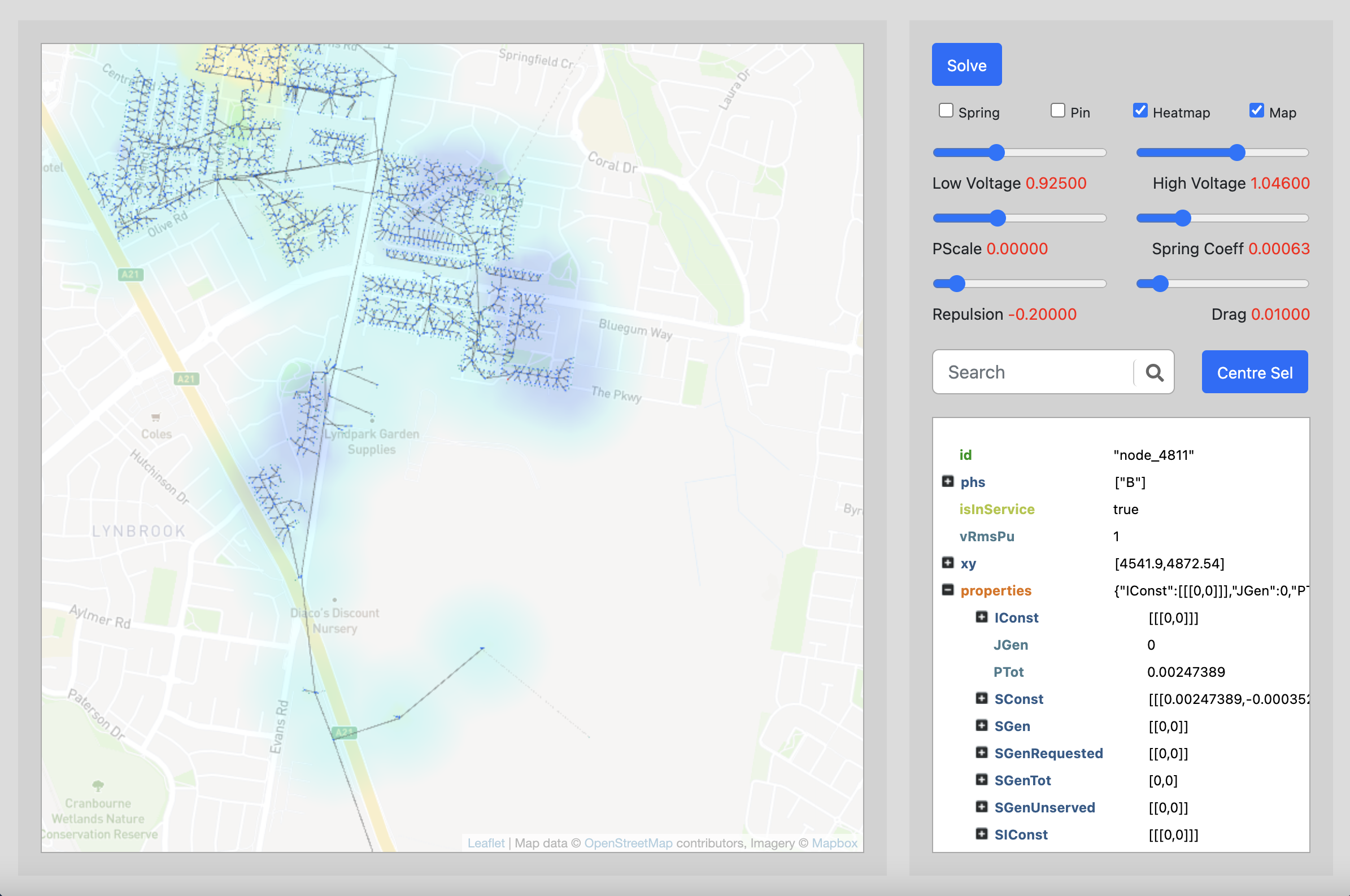}
    \end{center}
    \caption{
        Screenshot of the viewer.
    }
    \label{FIG_VIEWER}
\end{figure}

A JavaScript-based viewer and a backend web-server have been written for SmartGridToolbox, for visualizing networks. The web client can be used in a browser, and stand-alone cross-platform applications have also been written, using Electron \cite{electron}. The viewer allows the network to be displayed, optionally over a map. Spring layout can be performed, and voltages can be visualized using a heatmap. Network elements such as buses, loads, lines etc. are clickable, and properties of these objects can be displayed. Fig. \ref{FIG_VIEWER} shows a screenshot of the viewer.

\section{Case Studies}
In this section, we will look at case studies and examples that demonstrate some of the features of SmartGridToolbox in action. We will discuss two real-life field trials that have used SmartGridToolbox, and will follow this with a discussion of two example applications that are included with the SmartGridToolbox distribution.  

\subsection{CONSORT: Bruny Island Battery Trial} \label{SEC_CONSORT}
The CONSORT Bruny Island Battery Trial \cite{consort-webpage, consort-arena, thiebaux2019, franklin2016, scott2019, lovell2018, chapman2021, jurasovic2018} was a field trial run on Bruny Island, Tasmania, Australia, between 2016 and 2019. Involving three universities (The Australian National University, The University of Queensland and University of Tasmania), a distributed energy startup (Reposit Power) and a network utility (TasNetworks), the project successfully demonstrated a mean by which residential battery storage could be orchestrated in a near optimal way to alleviate network congestion and other issues while financially benefiting battery owners.

Bruny Island is a semi-rural island off the west coast of Tasmania. It is supplied with electricity by two undersea cables. The main cable is limited to about 64 A, and during peak holiday season, a diesel generator is used to supply current above this limit. Thus, it is a perfect setting for a small-scale study into how residential batteries can be used to alleviate network congestion.

At the heart of CONSORT was the distributed optimization algorithm, termed ``Network Aware Coordination'', or NAC \cite{scott2019}, that orchestrated home energy management systems supplied by Reposit Power. Using the ADMM algorithm \cite{boyd2011}, the NAC algorithm solved a distributed OPF and battery optimization problem, with the full problem being split into subproblems consisting of one OPF subproblem problem for each interval in a forward horizon, and one battery optimization subproblem for each participating household EMS, or ``remote''.

``NAC workers'' were responsible for solving the OPF subproblems, the EMS software, installed in ``Reposit Boxes'', was responsible for the battery optimization subproblems, and a ``NAC dealer'' was responsible for coordinating between the various subproblems according to the ADMM algorithm. The NAC dealer iteratively passed negotiated prices back and forth between the NAC workers and the EMSs (remotes), stopping when agreement was reached on the negotiated prices.

SmartGridToolbox was used for three main purposes throughout the project. Initially, it was used to investigate the general properties of the 11 kV network on the island, providing situational awareness.

Secondly, and more importantly, it was used as a form of ``acceptance testing'' prior to the deployment of the NAC algorithm on the actual physical network. It was during this phase that the advantages of using a flexible C++-based library were best demonstrated.

The NAC dealer and workers and the EMS optimization software were run in the cloud, with the latter being run in ``simulation'' mode under which the software was not hooked up to a real household, but was instead provided with all of its information via HTTP messages over the web.

The state of the network, all background loads, and the participating households (including PV and batteries) were simulated using SmartGridToolbox, with decisions about when to charge and discharge batteries being delegated to the cloud-based remotes (virtual Reposit boxes), based on information passed to them in messages by a custom \verb|SgtServer| SmartGridToolbox simulation component. This  \verb|SgtServer| component was responsible for mediating messages between the NAC dealer and the remotes, retaining relevant information for simulation purposes, and passing on additional simulation information about the state of batteries of participating households and predicted household usage and PV generation to the remotes.

\begin{figure}[htb]
    \begin{center}
        \includegraphics[width=\columnwidth-3cm]{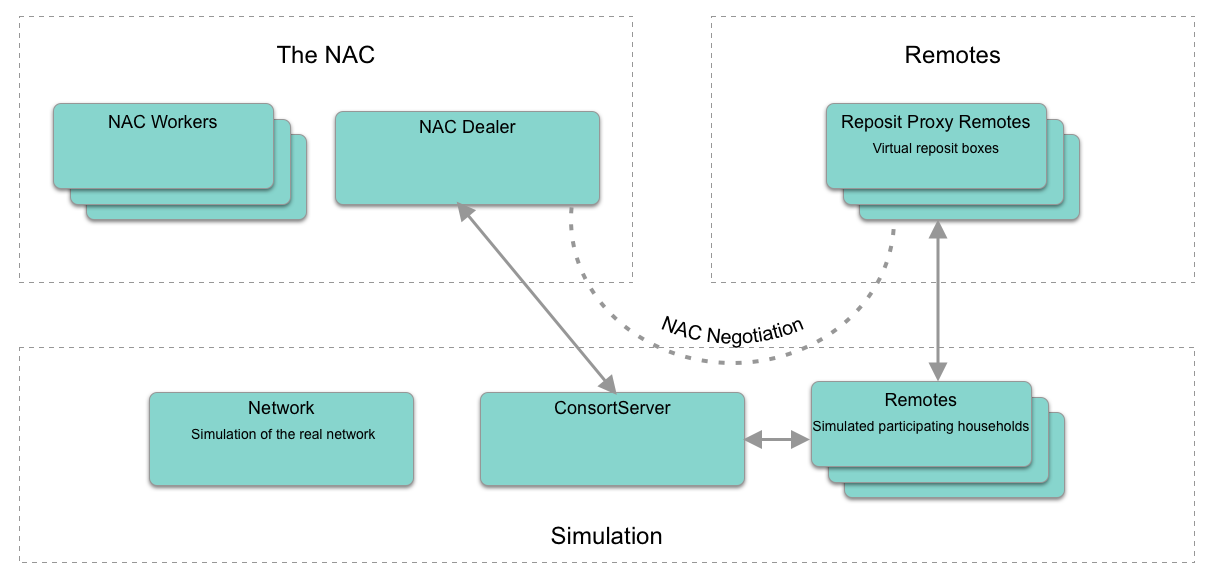}
    \end{center}
    \caption{
        Simulation architecture for the CONSORT trial.
    }
    \label{FIG_CONSORT_ARCH}
\end{figure}

A hybrid real-time/simulation-time approach was taken, where the simulation proceeded in real-time during NAC negotiations, which occurred every five minutes, and in simulation-time during periods when NAC was idle. The architecture is shown in Fig. \ref{FIG_CONSORT_ARCH}.

\begin{figure}[!t]
    \begin{center}
        \includegraphics[width=\columnwidth-5cm]{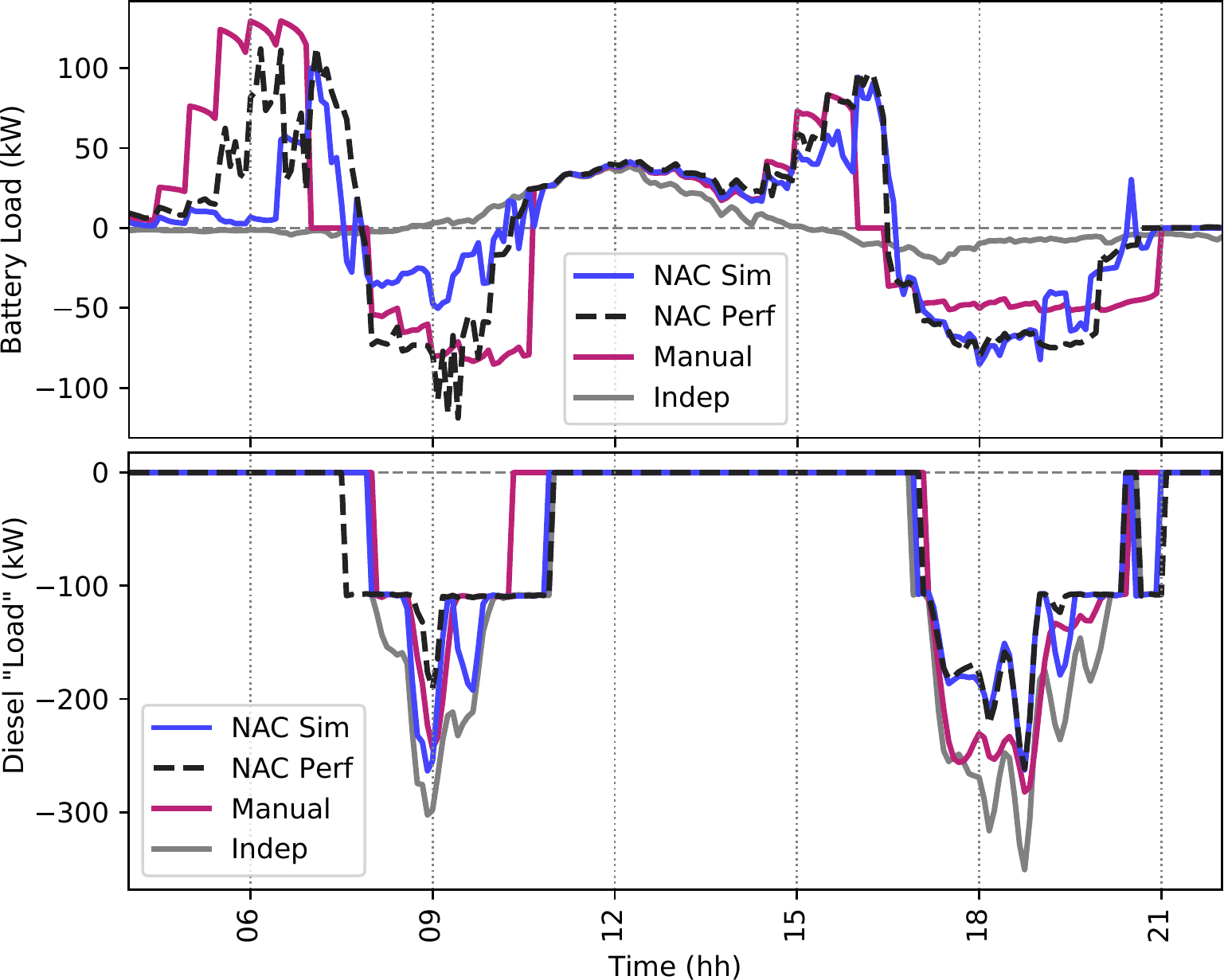}
    \end{center}
    \caption{
        Simulation of counterfactual ``what-if'' scenarios in the CONSORT Bruny Island Battery Trial, June 10, 2018. The top graph shows the combined load of participating batteries, and the bottom graph shows the load of the diesel generator. The grey curve shows what would happen if the batteries were acting without coordination and in their own self interest. This is to be contrasted with the blue curve, showing a realistic simulation of NAC with actual predictions for peak load, the black dashed curve, showing the improvement that can be gained by having perfect predictions of peak load, and the magenta curve, showing a typical manual aggregate dispatch of the batteries via a ``virtual power plant'' application.
    }
    \label{FIG_CONSORT_SIM}
\end{figure}

Finally, the same simulation setup described above was used to perform counterfactual analyses, to evaluate what-if questions. For example, this allowed us to assess the effectiveness of NAC in reducing the amount of energy provided by the diesel generator, by running the counterfactual scenario where the batteries acted independently, without the intervention of NAC. Fig \ref{FIG_CONSORT_SIM} shows a simulation of various counterfactuals during a long weekend on June 10, 2018. Using this kind of counterfactual approach, we were able to show that 31 participating households were able to reduce the use of backup diesel by around 34\% over the course of the trial \cite{scott2019}.

\subsection{The evolve DER Project}
SmartGridToolbox has been used for network modelling and solving in the evolve DER Project \cite{evolve-arena}, also funded by ARENA. The evolve project aims to increase the network hosting capacity of distributed energy resources (DER) while ensuring that network constraints are obeyed. It relies heavily on the concept of dynamic operating envelopes - a dynamic allocation of hosting capacity among DER \cite{arena-operating-envelopes}. SmartGridToolbox's python bindings are used to incorporate its network modelling and solving capabilities into evolve's workflow, with the AC power flow solver being used to help with the calculation of envelopes. 

\subsection{Volt-VAR control using PV inverters} \label{SEC_PV_DEMO}
In this demonstration, we consider the possibility of using residential solar PV inverters to manage voltage in networks where there is a high level of solar-PV penetration. The code can be found in the SmartGridToolbox source tree under \href{https://github.com/NICTA/SmartGridToolbox/tree/master/examples/PvDemo}{SmartGridToolbox/examples/PvDemo}. Solar PV may create voltage management issues, due to the rapid and unpredictable fluctuations in power that arise as clouds come and go. It has been suggested \cite{smyth2011} that the source of such problems may also form part of their solution: some solar PV inverters have the ability to act as controllable sources of reactive power, and are thus in principle able to perform volt-VAR control on the network. They have the advantage of being highly distributed throughout the network, and of not requiring any new sources of reactive power to be installed.

The example shown here uses the network of the IEEE 57 bus system \cite{ieee57}, which has been modified to create voltage issues in the network \cite{pglib} and then converted to \textsc{Matpower} format. Static loads are replaced by dynamic time series loads, which are created for each bus by randomly aggregating household load data. The maximum aggregate power at each bus is set to equal 1.2~times the static load from the ieee57 test case. The modified network is read into SmartGridToolbox from the YAML input file\footnote{See \url{https://gitlab.com/SmartGridToolbox/SmartGridToolbox/-/blob/master/examples/PvDemo/pvdemo_ieee57.yaml} for the full listing}:
\begin{lstlisting}
- matpower:
    sim_network_id: network
    input_file: data/ieee57/ieee57_0_mod.m
\end{lstlisting}
The dynamic loads are added using SmartGridToolbox's YAML looping construct:
\begin{lstlisting}[breaklines=true]
- loop:
    loop_variable: [i, 0, <n_ld_buses>, 1]
    loop_body: 
        - time_series: {id: load_<ld_buses(<i>)>_series, data_file: ...}
        - time_series_zip: {id: load_<ld_buses(<i>)>, time_series_id: ...}
\end{lstlisting}

Solar PV generation and special solar PV inverters are added using similar constructs
\begin{lstlisting}[breaklines=true]
- loop:
    loop_variable: [i, 0, <n_pv_buses>, 1]
    loop_body: 
        - pv_inverter: {id: pv_inv_<pv_buses(<i>)>, ...}
        - solar_pv: {id: solar_pv_<pv_buses(<i>)>, zenith_degrees: 60, ...}
\end{lstlisting}

In the listing above, the ``solar\_pv'' keyword creates a standard SmartGridToolbox \verb|SolarPv| component, which in this instance represents an aggregation of many individual rooftop systems. Likewise, the ``pv\_inverter'' keyword creates a custom \verb|PvInverter| component that was written for this example: a special inverter that also acts as a generator of reactive power. Each \verb|PvInverter| represents an aggregation of actual inverters. 
\begin{figure}[!h]
    \begin{center}
        \includegraphics[width=\columnwidth-2cm]{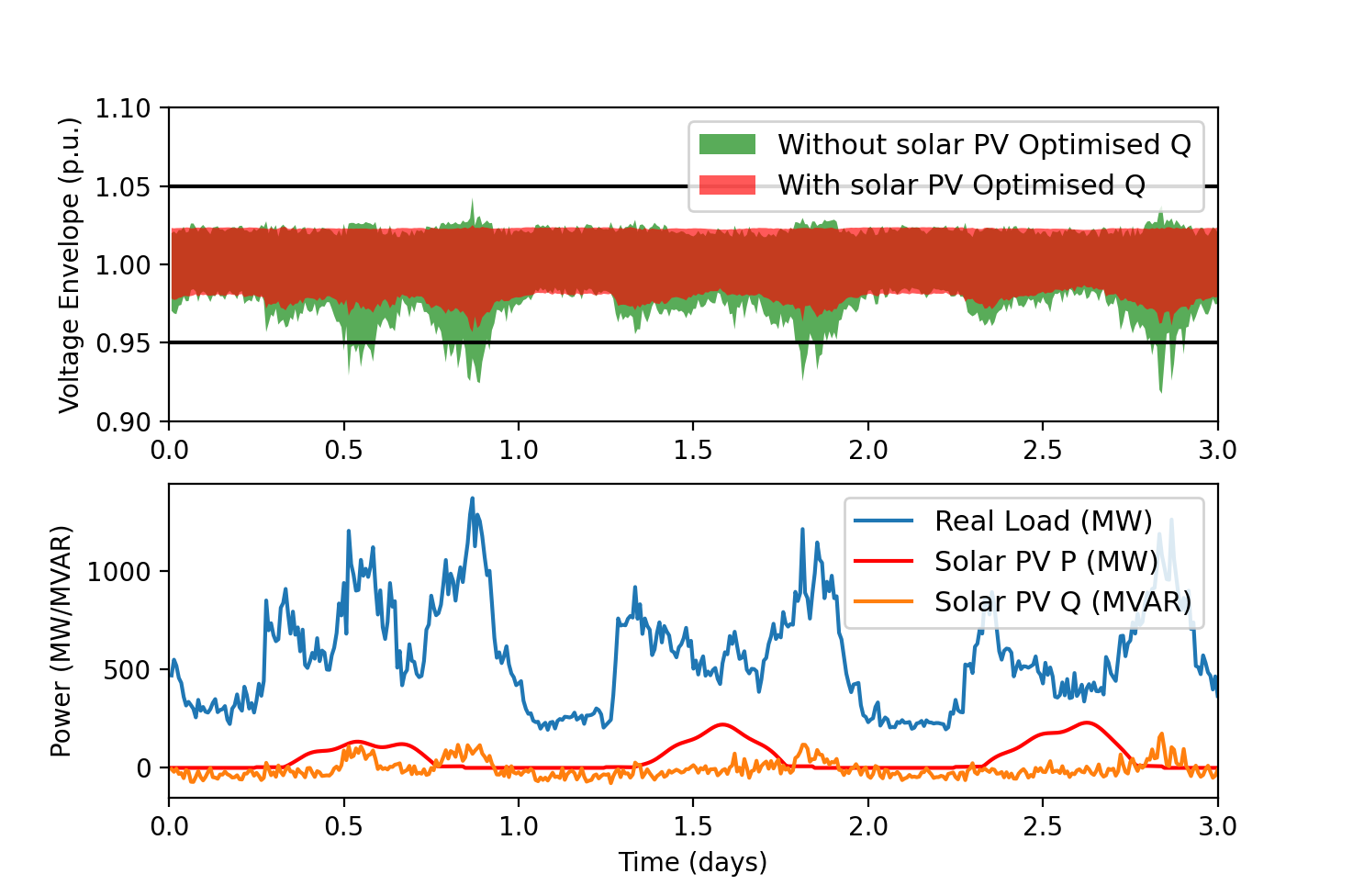}
    \end{center}
    \caption{
        Volt-VAR control using PV inverters. (Top) The voltage envelopes, in cases with optimized reactive generation by PV inverters (red), and without (green). Horizontal lines are the specified desired voltage limits. (Bottom) The load and real and reactive power from the PV inverters.
    }
    \label{FIG_PV_DEMO}
\end{figure}

The \verb|PvInverter| class works together with another customized class, \verb|PvDemoSolver|. \verb|PvDemoSolver| subclasses the standard OPF solver, discussed in section \ref{SEC_OPF_SOLVER}. Taking advantage of the extensibility of the OPF solver, extra variables and constraints are added: an additional penalized slack variable for voltage-bound violations (to handle cases where it is not possible or desirable to fully satisfy strict voltage bounds), a configurable penalty for deviations from nominal voltage, and extra constraints on the maximum apparent power of the \verb|PvInverter|s.

Results are shown in \ref{FIG_PV_DEMO}. The inverters are able to inject enough reactive power to keep the voltage envelope of all buses within the specified bounds. The control-case, on the other hand, has low voltage periods occurring during the afternoon/evening peaks. For the data shown, there is little or no trade-off between real power and reactive power at the solar inverters, as they are sized appropriately. However, we could also easily explore scenarios where some real power from the PV systems is sacrificed for reactive power, in conditions of unacceptable voltage peaks or drops.

\subsection{Optimized building control}\label{SEC_BUILDING_CONTROLLER}
\begin{figure}[htb]
    \begin{center}
        \includegraphics[width=\columnwidth-9cm]{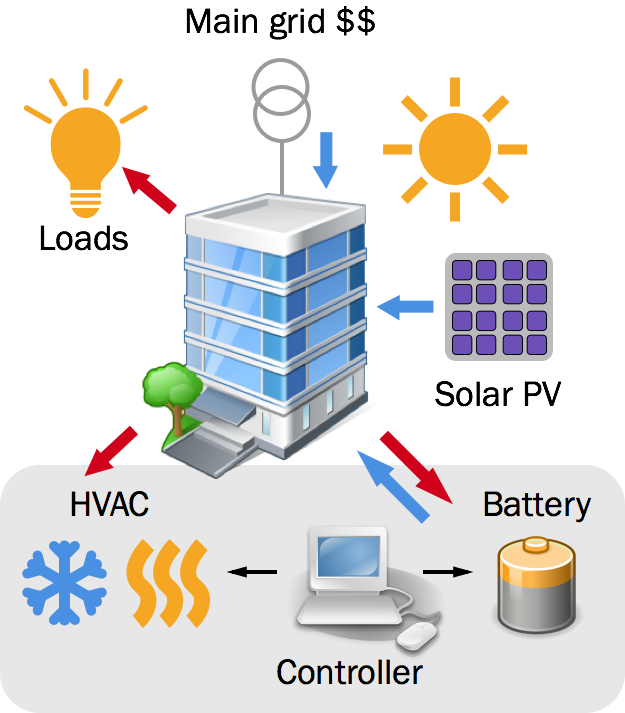}
    \end{center}
    \caption{
        Schematic of optimized building control. The HVAC and battery are optimized jointly to take advantage of fluctuations in the real time electricity price, external temperature, load profile and solar PV output.
    }
    \label{FIG_BUILDING_CONTROLLER}
\end{figure}
This example may be found in the SmartGridToolbox source tree under \href{https://github.com/NICTA/SmartGridToolbox/tree/master/examples/BuildingControllerDemo}{SmartGridToolbox/examples/BuildingControllerDemo}. Figure \ref{FIG_BUILDING_CONTROLLER} shows a schematic of the components in the example. A building buys power from the main grid at a fluctuating price, and may also sell back excess power at a fixed feed-in tariff that is lower than the retail price. It has solar PV, a controllable battery, a controllable HVAC system, and various uncontrollable loads. A building controller has access to predictions of electricity price, load profile, expected temperature and solar generation, and is able to charge and discharge the battery, and adjust the operation and comfort levels of the HVAC system in response to this information. For the purpose of this paper, we assume perfect predictions, but it is a simple matter to change the data to take into account imperfect knowledge. This example could also easily be modified to simulate optimized control of a microgrid containing multiple buildings, solar PV systems and batteries.

The controller performs a rolling horizon optimization every ten minutes. It discretizes time for the next 24 hours, and runs an optimization to find optimal charging/discharging levels of the battery and heating/cooling power for the HVAC system. The objective function is a sum of the cost of electricity purchased from the grid and a weighted deviation from a comfort band around the HVAC temperature setpoint. The optimal settings for the initial discretized timestep are used to update the controls on the battery and HVAC for the next ten minutes, with the rest of the optimized data being discarded. The linear optimization is implemented using the Gurobi optimizer \cite{gurobi} 

\begin{figure}[htb]
    \begin{center}
        \includegraphics[width=\columnwidth-5cm]{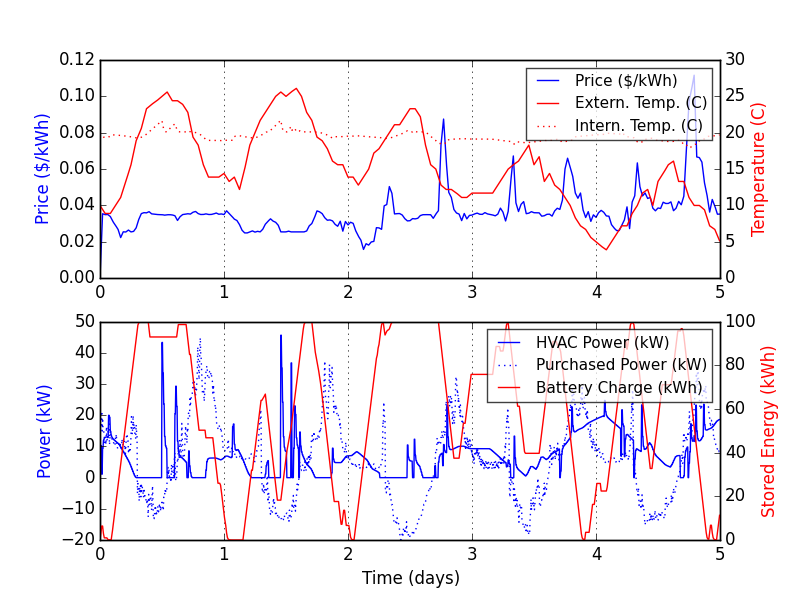}
    \end{center}
    \caption{
        Results for the optimized building example.
    }
    \label{FIG_BUILDING_CONTROLLER_RESULTS}
\end{figure}

Fig. \ref{FIG_BUILDING_CONTROLLER_RESULTS} shows some simulation results. The following features are apparent:
\begin{itemize}
    \item The battery discharges at times of high price and charges at times of low price.
    \item The battery may be undersized, in the sense that there are times when power is sold back to the main grid at a discount rather than being stored for later use.
    \item The HVAC clearly responds to external temperature fluctuations, heating and cooling as necessary.
    \item The internal temperature fluctuates around the comfort band, occasionally straying outside it as a trade-off against having to purchase power at a high price.
\end{itemize}

\section{Conclusion}
We have seen how SmartGridToolbox combines detailed, unbalanced network modelling with fast AC power flow and OPF solvers, and how the discrete event simulation part of SmartGridToolbox is built on top of this foundation. Several examples of applications of SmartGridToolbox have been presented, including it's significant role in two field trials.

Future additions may include distribution system state estimation, short-timescale dynamics, and sensitivity analysis (which is in fact already included in SmartGridToolbox, but is not yet well characterized).

It is our hope that SmartGridToolbox will continue to play a role in active projects, and that it will be of interest to a growing list of researchers and developers.

\section*{Acknowledgements}
SmartGridToolbox has greatly benefited from the opportunity to support two Australian field trials: the CONSORT Bruny Island Battery Trial and the evolve DER project. The Australian Government, through the Australian Renewable Energy Agency, provided \$2.9m towards the \$8m CONSORT project, and is providing \$4.29m towards the \$12.94m evolve DER project, under its Research and Development Program. The initial phase of development of SmartGridToolbox was assisted by funding from Actew-AGL.

\newpage
\printbibliography

\end{document}
